\documentclass[twoside,fleqn,epsfig]{article}
\usepackage{espcrc2}

\title{An upper limit to the electric dipole moment of the neutron
from lattice QCD}

\author{  B.~All\'es\address{INFN Sezione di Pisa,
                             Via Buonarroti 2, 56127-Pisa, Italy},
          M.~D'Elia\address{Dipartimento di Fisica,Universit\`a di Genova
                            and INFN Sezione di Genova,
                            Via Dodecaneso 33, 16146-Genova, Italy},
          A.~Di~Giacomo\thanks{Presented the talk.}\address{Dipartimento
                                di Fisica, Universit\`a di Pisa
                                and INFN Sezione di Pisa,
                                Via Buonarroti 2, 56127-Pisa, Italy}
}

\begin{document}

\begin{abstract}
A linear increase with the volume of the topological susceptibility
can signal spontaneous breaking of parity {\tt P}
and time inversion {\tt T}, due to a nonzero vacuum expectation value
of the topological charge $Q$. Such a breaking would produce a nonzero
electric dipole moment of the neutron, $d_n$. An upper limit to $d_n$
is derived from numerical simulations at increasing volumes.
\end{abstract}

\maketitle

\section{INTRODUCTION}

The QCD lagrangian
\begin{eqnarray}
{\cal L}_{\rm QCD} &=& {1\over 2} 
{\rm Tr}\{G_{\mu\nu}G_{\mu\nu}\} + \nonumber\\
&&  \sum_f \overline{\psi}_f \left(i{\rlap/D} - m_f\right)\psi_f
\end{eqnarray}
is invariant under parity {\tt P} and time
inversion transformations {\tt T}. Adding
to ${\cal L}_{\rm QCD}$ a Chern--Simons term
\begin{equation}
{\cal L}_{\rm CS}\equiv\theta Q(x) \, ,
\end{equation}
with $Q(x)$ the topological charge density
\begin{equation}
Q(x)={g^2\over 64\pi^2}\epsilon_{\mu\nu\rho\sigma}{\rm Tr}
\{G_{\mu\nu}G_{\rho\sigma}\}
\end{equation}
does not modify the equations of motion since $Q(x)$ is a
divergence, $Q(x)=\partial_\mu K_\mu(x)$,
\begin{equation}
K_\mu={g^2\over 16\pi^2}\epsilon_{\mu\nu\rho\sigma}
A^a_\nu\left(\partial_\rho A^a_\sigma - {1\over 3} g
f^{abc} A^b_\rho A^c_\sigma\right)\,.
\end{equation}
The lagrangian ${\cal L}_{\rm CS}$ has a dynamical effect,
being sensitive to the boundary conditions, and its presence
in $\cal{L}\equiv {\cal L}_{\rm QCD} + {\cal L}_{\rm CS}$
results in an explicit breaking of {\tt P} and {\tt T}.
Indeed, since $Q\sim \vec{E}\cdot\vec{H}$ it is odd
under both of them. This breaking can produce a nonzero
value of $d_n$, the electric dipole moment of the 
neutron~\cite{baluni}, $d_n\approx8\cdot 10^{-16}\theta {\rm e}\cdot
{\rm cm}$.
The experimental upper limit is~\cite{expt} $d_n|_{\rm exp}< 6\cdot 10^{-26}
{\rm e}\cdot {\rm cm}$ which implies
$\theta {\buildrel < \over \sim} 10^{-10}$.
This is a small number and its smallness is not protected by
any known symmetry.

An alternative possibility for having $d_n\not= 0$ could be a
spontaneous breaking of {\tt P} by a nonzero value
$\left\langle 0 | Q | 0 \right\rangle$. This possibility however
would be excluded by a theorem by Vafa and Witten~\cite{vafawitten}.
Nevertheless this theorem is based on assumptions whose validity
has been questioned. We have then addressed the issue of a spontaneous
breaking of {\tt P} and {\tt T} by numerical simulations of QCD
on the lattice~\cite{alles}.

\section{THE VAFA--WITTEN THEOREM}

The statement is that in a theory of gauge fields and fermions
like QCD which is Lorentz invariant and with vector interactions,
the v.e.v. of any operator $O$ odd under parity is zero. The argument
goes schematically as follows: let us consider the generalized
partition function
\begin{equation}
Z[\theta ] = \int {\cal D}\varphi \;\;{\rm e}^{-i\left( S(\varphi) +
 \theta\int O {\rm d}^4x\right)} \,,
\end{equation}
where $\varphi$ represents the generic field variables,
and its continuation to the euclidean
\begin{equation}
Z_{\rm Euclid}[\theta]\equiv {\rm e}^{-V E(\theta)}\, ,
\end{equation}
where $E(\theta)$ is the free energy as a function of the
$\theta$ parameter and $V$ is the spacetime volume.

After the analytic continuation the measure is positive since the
action is real and the fermion determinant is positive. The
operator $O$ instead acquires an extra factor $i$ since, to be
odd under {\tt P} and {\tt T}, it must have an odd number of
invariant tensors $\epsilon_{\mu\nu\rho\sigma}$, at least if
it depends only on the gauge fields like $Q(x)$.

As a consequence the following equality holds
\begin{equation}
{\rm e}^{-V E(\theta)} =\int {\cal D}\varphi \;\;{\rm e}^{- S(\varphi) - i
 \theta\int O {\rm d}^4x}\, .
\end{equation}
If $O$ is hermitean then the term $\exp\left(-i\theta\int O {\rm d}^4x\right)$
becomes a phase factor and since the rest of the integrand is positive,
the inequality
\begin{equation}
{\rm e}^{-V E(\theta)} \leq {\rm e}^{-V E( 0 )}
\end{equation}
derives. Consequently we have $E(\theta)\geq E(0)$, i.e. $\theta=0$
is a minimum. Since for small $\theta$, $E(\theta)\approx 1 - \theta 
\left\langle 0 | Q | 0 \right\rangle$, it implies
$\left\langle 0 | Q | 0 \right\rangle=0$.

There are a number of assumptions behind this result.

\begin{itemize}
\item{} Lorentz invariance, which is certainly true for QCD at
zero temperature, is lost at finite temperature. Then, the
breaking of parity, $\left\langle 0 | Q | 0 \right\rangle\not=0$
becomes plausible~\cite{cohen,kharzeev} with possible observable
consequences in heavy ion collisions.

\item{} The argument can be nonvalid if $O$ is constructed
not only by use of gauge fields but also 
with fermionic fields~\cite{aokisharpe}.

\item{} Analyticity in $\theta$ of $Z[\theta ]$ is assumed.
This hypothesis has been debated in the 
literature~\cite{azcoiti,aguadoasorey,asorey} and it is
in fact the one which we have tried to test on the lattice for the
topological charge operator.

\end{itemize}

Suppose indeed that $E(\theta )$ has a minimum at $\theta=0$, but a
discontinuous derivative (a cusp) at this point so that
\begin{equation}
 \frac{{\rm d} E(\theta)}{{\rm d}\theta}|_{\theta=0^\pm} =
 \pm \alpha\, ,
\end{equation}
or
\begin{equation}
\langle 0 \vert  Q \vert 0 \rangle\vert_{\theta=0^\pm} = \pm V\alpha\,.
\end{equation}
Then a spontaneous breaking of {\tt P} and {\tt T} would result.

The topological susceptibility
\begin{equation}
\chi = \int {\rm d}^4x\; \langle 0 \vert  Q(x) Q(0) \vert 0 \rangle =
\frac{\langle 0 \vert  Q^2 \vert 0 \rangle}{V}
\label{chi}
\end{equation}
would then contain the usual contribution coming from
vacuum fluctuations (which in the present paper shall be called
$\overline{\chi}$), and another contribution derived from
the nonzero vacuum expectation $\langle 0 \vert  Q \vert 0 \rangle$,
\begin{equation}
\chi = \alpha^2 V + \overline{\chi}
\end{equation}
and $\chi$ would be infrared divergent.

The electric dipole moment of the neutron $d_n$ can be estimated
as in~\cite{baluni}. It is proportional to the parameter
$\alpha$ times a typical volume $m_n^{-4}$ times a factor $\left(
m_\pi/m_n\right)^2$ which makes it vanish in the chiral limit,
\begin{equation}
d_n = e \frac{\alpha}{m_n^4} \frac{m_\pi^2}{m_n^3} \, .
\label{dn}
\end{equation}
An extra factor $1/m_n$ is needed to assign the correct length units.
{}From lattice we will get an upper limit for $\alpha$ and, from
Eq.~(\ref{dn}), an upper limit for $d_n$.

\section{LATTICE ANALYSIS}

Lattice is an UV regulator of the theory. The topological charge $Q$
can be determined either directly for the discretized version $Q_L$
of the charge operator (the so--called field--theoretical method) or from
the counting of zero modes in the fermionic degrees of freedom~\cite{jan}
by using the anomaly equation
\begin{equation}
\partial_\mu J_{5\;\mu} = 2 N_f Q(x) \,.
\end{equation}
The latter method requires a special discretization of the 
fermion fields~\cite{gw}
which at the present state of the art is rather demanding
in computer time, especially when one has to deal with large
volumes as in our case.

In the quenched theory the general relations
\begin{equation}
Q_L=Z Q\,, \qquad \chi_L\equiv\frac{\langle Q_L^2\rangle}{V} =
 Z^2 a^4 \chi + M
\end{equation}
hold. $Z$ is a multiplicative constant~\cite{vicari}
that can be determined by the expression
\begin{equation}
Z= \frac{1}{n}\langle Q_L\rangle\vert_{Q=n}
\end{equation}
where the v.e.v. is calculated in the topological charge sector
$Q=n$ where $n$ is a nonzero integer. The procedure of selection
of the set of configurations belonging to this sector is performed
by cooling. On the other hand $M$ is an additive renormalization
due to singularities that arise after the limit $x\rightarrow 0$
in the integrand of Eq.~(\ref{chi}). $M$ is an UV effect that can
be singled out by calculating $\chi_L$ in the zero charge topological
sector, $M=\chi_L\vert_{Q=0}$~\cite{gunduc}

The Ginsparg--Wilson formalism allows to preserve exact chiral
symmetry in the theory. In this case $Z=1$ and $M=0$~\cite{venegiusti}.
In the field-\break theoretical method these renormalization constants
must be calculated with the help of cool-\break 
ing--heating methods~\cite{campos,gunduc}.

We have adopted the field--theoretical method. The simulations
have been done on volumes
$16^4$, $32^4$ and $48^4$ at $\beta=6/g^2_{\rm bare}=6.0$,
i.e. at lattice spacing $a\approx 0.1$ Fermi. The statistics
was $12\cdot 10^4$, $6\cdot 10^4$ and $5\cdot 10^4$ independent measurements
for the three volumes respectively.

We have first determined the dependence of $Z$ and $M$ on
the lattice size $L=V^{1/4}$. In
Figures~1 and~2 we show the corresponding results.
Only the additive renormalization $M$, which is an extensive
quantity, displays a clear size dependence.
\begin{figure}[htb]
\vspace{4.5cm}
\includegraphics{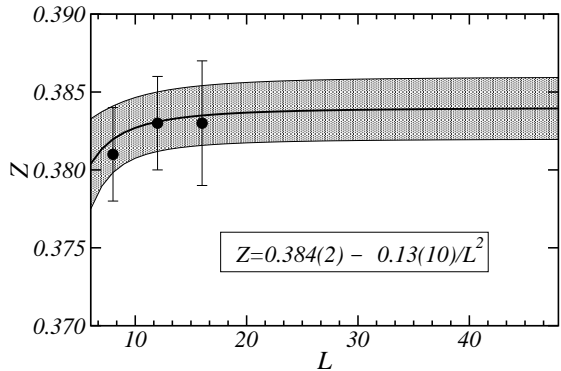}
\null\vskip 0.3cm
\caption{$Z$ versus $L$ for the lattice definition of the topological
charge operator $Q_L$ adopted in the present paper at $\beta=6.0$.
The line is the result of the fit (shown in the legend) and the grey band
is the 1--$\sigma$ error.}
\label{fig:Fig1}
\end{figure}
\begin{figure}[htb]
\vspace{4.5cm}
\includegraphics{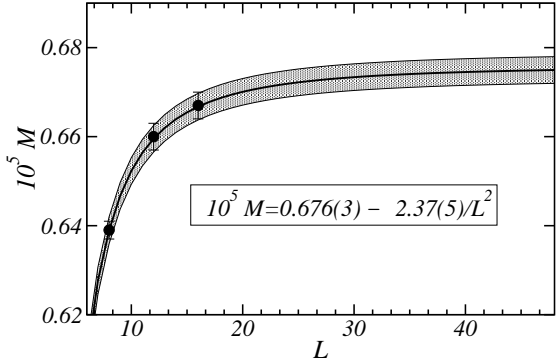}
\null\vskip 0.3cm
\caption{$M$ versus $L$ for the lattice definition of the topological
susceptibility $\chi_L$ adopted in the present paper at $\beta=6.0$.
The line is the result of the fit (shown in the legend) and the grey band
is the 1--$\sigma$ error.}
\label{fig:Fig2}
\end{figure}
Then, knowing $Z$, $M$ and the lattice spacing $a$ we have extracted
$\chi$ as a function of the volume $V=L^4$. The result is shown
in Figure~3.
\begin{figure}[htb]
\vspace{4.5cm}
\includegraphics{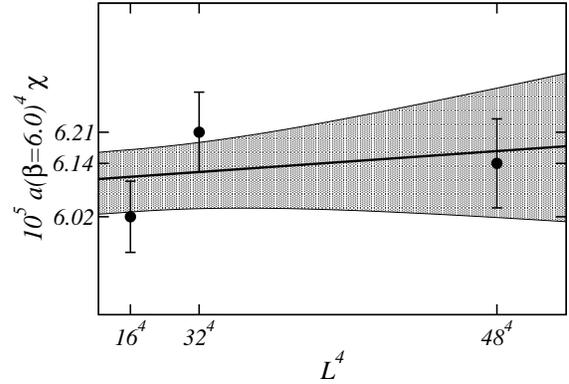}
\null\vskip 0.3cm
\caption{$a^4\,\chi$ versus $L^4$ at $\beta=6.0$. The straight line
and grey band are the result of the fit and its 1--$\sigma$ error
respectively.}
\label{fig:Fig3}
\end{figure}

No linear dependence on $L^4$ is found within errors and this allows
to put the upper bound (within 1--$\sigma$)
\begin{equation}
\alpha \leq\left(\frac{1}{4 \;{\rm Fermi}}\right)^4\, ,
\end{equation}
or, by use of Eq.~(\ref{dn}),
\begin{equation}
d_n\leq 3.5 \; 10^{-21} {\rm e}\cdot{\rm cm}\, .
\end{equation}
As a byproduct, we found that the physical topological
susceptibility at $a\approx 0.1$ Fermi
$\overline{\chi}$ (or $\chi$ if $\alpha$ vanishes) is
$\left(173.4(\pm 0.5)(\pm 1.2)(^{+1.1}_{-0.2})\;{\rm MeV}\right)^4$ 
where the errors are the statistical error from our data, the
one derived from the value used for $\Lambda_L$ and an estimate of
the systematic error respectively. This result agrees with previous
determinations~\cite{ADD} and with the Witten--Veneziano formula~\cite{witten}
needed to explain the large $\eta'$ mass.

\section{CONCLUDING REMARKS}

Our upper limit on $d_n$ is 5 orders of magnitude bigger than the experimental
one. It can, however, be improved by increasing the lattice size and by
decreasing the statistical errors.

We are repeating the computation at nonzero temperature to check the
ideas of reference~\cite{kharzeev}. The field--theoretical method
to calculate the topological susceptibility on the lattice is fast and
efficient and is presently the only one which can afford large
lattices.

The calculation was done on the APEmille facility of INFN in Pisa.

This work is partially financed by MIUR, Program 
``Frontier Problems in the Theory of Fundamental Interactions''.

\end{document}